\documentclass[graybox, envcountchap]{svmult}

\usepackage{mathptmx}        
\usepackage{helvet}          
\usepackage{courier}         

\usepackage{makeidx}         
\usepackage{graphicx}        
\usepackage{multicol}        
\usepackage[bottom]{footmisc}

\makeindex             

\usepackage{amsmath}
\usepackage{amssymb}
\usepackage{lscape}
\usepackage{multirow}

\def\aapr{A\&A Rev. }
\def\aap{A\&A}%
\def\aj{AJ }
\def\apjl{ApJL }
\def\apjs{ApJS }
\def\apj{ApJ }
\def\apss{Ap\&SS}%

\def\gapp{\ifmmode\stackrel{>}{_{\sim}}\else$\stackrel{<}{_{\sim}}$\fi}
\def\gsim{\lower.5ex\hbox{\gtsima}}
\def\gtsima{$\; \buildrel > \over \sim \;$}
\def\lapp{\ifmmode\stackrel{<}{_{\sim}}\else$\stackrel{<}{_{\sim}}$\fi}
\def\lsim{\lower.5ex\hbox{\ltsima}}
\def\ltsima{$\; \buildrel < \over \sim \;$}

\def\mnras{MNRAS }
\def\pasj{PASJ}%

\def\zap{ZAp}%

\newcommand\apgt{\ {\raise-.5ex\hbox{$\buildrel>\over\sim$}}\ }
\newcommand\aplt{\ {\raise-.5ex\hbox{$\buildrel<\over\sim$}}\ }
\newcommand{\ga}         {\gtrsim}%

\newcommand{\la}         {\lesssim}

\begin{document}
\pagestyle{empty}
\frontmatter

\include{dedic}
\include{foreword}
\include{preface}

\mainmatter

\setcounter{chapter}{7}
\title{Binary Evolution: Roche Lobe Overflow and Blue Stragglers}\index{Roche lobe overflow}
\author{Natalia Ivanova}
\institute{Natalia Ivanova \at University of Alberta, Edmonton, AB T6G 2G7, Canada, \email{nata.ivanova@ualberta.ca}
}
%
%
\maketitle
\label{Chapter:Ivanova}

\abstract*{One of the principal mechanisms that is responsible for the origin of
blue stragglers
is  mass transfer that takes place while one of the binary companions
overfills its Roche lobe.
In this Chapter, we overview the theoretical understanding of mass
transfer via Roche lobe overflow: classification,
how both the donor and of the accretor respond to the mass transfer on
different timescales (adiabatic response, equilibrium response,
superadiabatic response, time-dependent response) for different types
of their envelopes (convective and radiative).
These responses, as well as the assumption on how liberal the process
is, are discussed
in terms of the stability of the ensuing mass transfer.
The predictions of the  theory of mass transfer via Roche lobe
overflow are then briefly compared
with the observed mass-transferring systems with both degenerate and
non-degenerate donors.
We conclude with the discussion which cases of mass transfer and which
primordial binaries
could be responsible for blue stragglers formation via Roche lobe
overflow, as well as how this can be enhanced
for blue stragglers formed in globular clusters}

%
%
%

%
%
%
%

\section{Introduction}

As we have seen in Chap. 7, when a star (the donor) in a binary system
has a radius larger than some critical value, called the Roche lobe\index{Roche lobe},
it will transfer mass\index{mass transfer} to its companion (the accretor), through Roche lobe overflow\index{Roche lobe overflow} (RLOF).
Both the beginning of the RLOF and its outcome 
depend strongly on
what {\it kind} of a donor\index{donor star} had started the mass transfer. 
Several characteristics are commonly considered to be most important 
for predicting the outcome: 

\begin{enumerate}
\item the evolutionary status of the donor -- this implies its 
complete internal structure;
\item the structure of the donor's envelope;
\item the mass ratio\index{mass ratio} of the binary;
\item the type of the accretor\index{accreting star}
\end{enumerate}

\noindent All these properties are intrinsically responsible 
for the stability of the mass transfer and for its final product.

As opposed to those governing features, the mass transfer can also be discussed in terms of its case.
The {\bf case} of the mass transfer is simply a {\it classification} of the mass transfer 
by the evolutionary status of the donor as the following:

\begin{itemize} 
\item {\bf Case A}\index{Cases A, B, C of mass transfer} -- during hydrogen (H) burning\index{hydrogen burning} in the core of the donor. 
\item {\bf Case B} -- after exhaustion of hydrogen in the center of the donor.
\item {\bf Case C} -- after exhaustion of central helium (He) burning\index{helium burning}.
\end{itemize}

\noindent  The case A and the case B were first introduced by Kippenhahn \& Weigert \cite{kipwei67}, 
and the case C was later introduced by Lauterborn \cite{ivalau70}
(he cited \cite{kipwei67} for the compete A-B-C classification, 
however no traces of the case C discussion there could be found).
Interesting, that when the case C was introduced, 
the specific clarification for the case B was made that
the central He burning has not yet started. 
This left the central He core burning\index{core burning} in limbo for the purpose 
of this simple classification, 
and in literature it has been filed variably as cases B or C. 

Nonetheless, it is important to realise that the specific case of the mass transfer 
by itself does not imply whether the mass transfer would be stable or not, 
as it was based on the current nuclear energy source.
The evolutionary stage and internal structure\index{internal structure} of the donor are more 
important in determining 
what object could be left at the end
of the mass transfer\index{mass transfer} {\it if} the initial stability and the timescale of the initiated mass transfer are known by other means. 
The structure of the inner layers may affect the stability of the mass 
transfer later on, for example, when the 
initial core becomes exposed   (for further details see \S\ref{sec_ddi}).
It is the structure of the donor's envelope that affects the initial stability.
 
\begin{figure}
\sidecaption
\includegraphics[width=75mm]{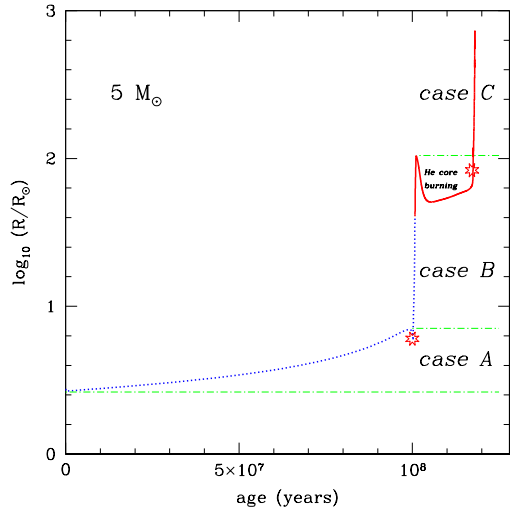} 
\caption{Evolution of the radius of a single star of 5 M$_\odot$ (metallicity $Z = 0.02$). 
The dotted (blue) line indicates that the star has radiative 
envelope and solid (red) line indicates that the star 
has a convective envelope. Two ``star'' symbols indicate when 
hydrogen is exhausted in the core and when helium is exhausted in the core.
Green dash-dotted lines separate the cases of the mass transfer 
in an assumption that a binary does not shrink.}
\label{mtcase}
\end{figure}

Any of the cases of the mass transfer could start with any type of donor's envelope -- 
either radiative\index{radiative envelope} or convective envelope\index{convective envelope} (see Fig.~\ref{mtcase} 
for a typical radius evolution), 
especially considering that case C {\it might} include the He core burning stage.
On the other hand, Roche lobe overflow (RLOF) might not start 
at some of evolutionary stages.
For instance, unless the binary has been perturbed so as to shrink,
a donor usually can not have RLOF if its radius at this moment
is smaller than at any moment of the donor's life before (otherwise it would have undergone the RLOF earlier). 
To alter that, an angular momentum loss from the binary needs to occur. 
While the donor evolves, its timescales for the next evolutionary stages 
speed up compared to main sequence lifetime.
Coupled with the donor's wind mass loss\index{wind mass loss}, this usually prevents the 
start of the mass transfer if a potential  donor
has contracted during, e.g., its He core burning (see Fig.~\ref{mtcase}). 

\section{Stability of the Mass Transfer: The Global Condition}

When a donor starts it mass transfer, two things happen at the same time. 
First, a mass-loosing star changes its radius.
The response of the donor radius $R_{\rm d}$ to the mass-loss can be written 
as $R_{\rm d}\propto M^{\zeta_*}$. 
During the mass transfer, bits of the donor's material either move from one star to another, 
or they are lost from the system. Angular momentum then
could be transferred to a companion, or to a circumbinary disc, 
or again it could be lost from the binary system 
via magnetic braking, or gravitational wave radiation, or simply with the lost material.
The Roche lobe's response to the mass loss can be written  as $R_{\rm L} \propto M^{\zeta_{\rm L}}$. 
These two responses are known as mass-radius exponents and were first introduced in \cite{webbink1985} (see also \cite{Hje87} for more details).

\begin{eqnarray}
\zeta_{\rm L}\equiv  \frac{d \log R_{\rm L}}{d \log M}\ .
\label{zetal}
\end{eqnarray}

\begin{eqnarray}
\zeta_*\equiv \left ( \frac{\partial \log R_{\rm d}}{\partial \log M} \right )\ .
\end{eqnarray}
\noindent In the last exponent, the changes of the donor radius are only due to its evolution and the mass loss.

In order to continue the mass transfer {\it stably}, the donor must remain within its Roche lobe. 
Hence, the stability of the mass transfer depends on the donor's 
response to the mass loss (this is implicit in $\zeta_*$) and on how conservative the process 
is and what are angular momentum loss processes in this binary 
(this is implicit in $\zeta_{\rm L}$). 
Then 
\begin{eqnarray} 
\zeta_{*} \ge \zeta_{\rm L}  & \Longleftrightarrow \mbox{\ \ stability of the mass transfer ;} \ \ \nonumber \\
\zeta_{*} < \zeta_{\rm L} & \Longleftrightarrow \mbox{ instability of the mass transfer} . 
\end{eqnarray}

\noindent The larger $\zeta_*$,  the more stable the mass transfer is.

This condition -- due to the donor response -- represents only one of the possible situations leading to the unstable mass transfer.
Unstable mass transfer can also be a result of the {\it accretor's} response on a too fast mass transfer. 
For example, too fast accretion rate can reincarnate a white dwarf into a red giant \cite{Nom79}, or build up of a ``trapped'' envelope around a neutron star during a super-Eddington mass-transfer \cite{Beg79}. More discussion on these cases can be found in \cite{iva_ce}. 
Here we will only provide an example that is more applicable for blue straggler formation: 
an accretor can overfill its own Roche lobe, due to its evolution, while mass transfer from the primary
is still continuing.
When the spin angular momentum of the system is more than a third of its orbital angular momentum, the system is secularly unstable with respect to mass transfer. 
This is  known as the the Darwin instability \cite{dar79,Hut80}. 

\section{Roche Lobe Response}

As mentioned in Chap. 7, the Roche lobe radius is given by \cite{Eggl83}:

\begin{equation}
\frac{R_{\rm L}}{a} = \frac{0.49 q^{2/3}}{0.6 q^{2/3}+\ln (1+q^{1/3})}\ .
\end{equation}

\noindent Here $q$ is the mass ratio ($q=M_{\rm d}/M_{\rm a}$, where $M_{\rm a}$ is the accretor mass) and $a$ is the binary's orbital separation. 
For the useful range of mass ratios $0.1<q<10$ , this Roche lobe radius also can be further approximated as \cite{Eggl06}:

\begin{equation}
\frac{R_{\rm L}}{a} \approx 0.44 \frac{q^{0.33}}{(1+q)^{0.2}}\ .
\end{equation}

\noindent The response of the Roche lobe in the case when the mass transfer is fully conservative -- neither mass nor angular momentum is lost from the binary --  is then \cite{Tout97}

\begin{equation}
\zeta_{\rm L}=2.13 q-1.67 \ .
\label{zetal_cons}
\end{equation}
The mass transfer will become unstable if the mass ratio is higher than the critical value:

\begin{equation}
q > q_{\rm crit} = \frac{\zeta_*}{2.13} + 0.788 \ .
\end{equation}

\noindent $q_{\rm crit}$ provides a useful way to estimate roughly whether the mass transfer\index{stable mass transfer} would be stable or unstable\index{unstable mass transfer}.
We emphasise, however, that the intrinsic condition of dynamically unstable mass transfer is given by the donor response, $\zeta_*$, 
and is only expressed via $q_{\rm crit}$. This is frequently forgotten and $q_{\rm crit}$ is used at face value only, despite the fact that the donor's response, 
even for the same class of donors, varies.

As opposed to \emph{conservative} mass transfer\index{conservative mass transfer}, a mass transfer evolution where mass or angular momentum is taken away from the system
and, hence, is not conserved has been named {\it liberal}\index{liberal mass transfer} \cite{2000NewAR..44..111E}.
A great amount of details on how $\zeta_{\rm L}$ changes in cases of various modes of mass and angular 
momentum loss can be found in 
\cite{Sob97}. We only consider here the next most important case, when only a fraction of the transferred 
material, $\beta$, is accreted onto $M_{\rm a}$ 
(with $\beta=1$ implying that the mass transfer is fully conservative). 
Eq.(\ref{zetal}) expands as 

\begin{equation}
\zeta _{L} =  \frac{\partial \ln a}{\partial \ln m_{\rm d}} + \frac{\partial \ln (R_{L}/a)}{\partial \ln q}\frac{\partial \ln q}{\partial \ln m_{\rm d}}\ .
\end{equation}

\noindent Here $\partial \ln a/ \partial \ln m_{\rm }$ is solely due to the mass loss or transfer: 

\begin{equation}
\frac{\partial \ln a}{\partial \ln m_{\rm d}} =  \frac{m_{\rm a}}{\dot m_{\rm d}} \frac{\dot a}{a}
 =  \frac{ 2 q^2 - 2  - q (1-\beta)}{q+1}\ .
\label{nonc_eq1}
\end{equation}
\noindent The second term consists of the mass ratio response to the change in donor mass,

\begin{equation}
\frac{\partial \ln q}{\partial \ln m_{\rm d}} = 1 + \beta q ,
\label{nonc_eq2}
\end{equation}
\noindent and the Roche lobe response to the change in mass ratio, which can be found  
using Eggleton's approximation \cite{Eggl83}:

\begin{equation}
\frac{\partial \ln (R_{L}/a)}{\partial \ln q} = \frac{2}{3} - \frac{q^{1/3}}{3}\frac{1.2q^{1/3} + 1/(1 + q^{1/3}) }{0.6q^{2/3} + \ln (1 + q^{1/3})}\ .
\end{equation}

\noindent The derivations above are important to  show that if the mass transfer is becoming non-conservative ($\beta < 1$), the value of $\zeta_{\rm L}$ is decreasing. This implies that non-conservative mass transfer is stable over a larger range of mass ratios than conservative mass transfer.

\section{Donor's Response}

\subsection{Timescales}

If a star suddenly loses mass, then both hydrostatic\index{hydrostatic equilibrium} and thermal equilibria\index{thermal equilibrium} are disturbed, and a star
starts to readjust its structure in order to recover both equilibria.
The readjustment of the internal structure results in the star's radius evolution 
and, hence, can be described using $\zeta_*$. 
 
\vskip0.2cm

\noindent{\bf Hydrostatic readjustment} 
The donor's dynamical timescale is usually much smaller than its thermal timescale, $\tau_{\rm dyn }\ll \tau_{\rm KH}$.
Therefore, the initial donor's response to mass loss can be expected to be almost adiabatic.
Here we clarify that the term ``adiabatic'' in this context means that the 
entropy of each mass shell of the donor remains the same as at the start of the mass loss, as they are considered to have no time to exchange energy
with their neighbour layers -- in other words, the entropy profile of the donor is frozen (note the difference with the usual use of the term adiabatic, where it implies that the entropy of the whole systems is conserved).
This response for the immediate stability on a dynamical timescale is known as the adiabatic mass-radius exponent \cite{webbink1985}:

\begin{equation}
\zeta_{\rm ad}\equiv \left (\frac{\partial \log R_{\rm d}}{\partial \log M} \right )_{\rm ad} \ \ \Rightarrow \ \ 
\mbox{if  } \zeta_{\rm L} \le \zeta_{\rm ad}\mbox{, \ the mass transfer is {dynamically stable}}.
\label{zetaad}
\end{equation}

\noindent The mass transfer is dynamically stable if the donor shrinks within its Roche lobe on $\tau_{\rm dyn}$. 

\vskip0.2cm

\noindent{\bf Thermal readjustment.}
At the same time, the donor  will attempt to recover its thermal equilibrium, 
readjusting its internal structure to obtain a radius that is appropriate  for its new decreased mass.
This response is known as thermal-equilibrium mass-radius exponent \cite{webbink1985}:

\begin{equation}
\zeta_{\rm eq}\equiv \left (\frac{\partial \log R_{\rm d}}{\partial \log M} \right )_{\rm eq}  \ \ \Rightarrow   
\left\{
 \begin{array}{l l} 
\mbox{if  } \zeta_{\rm eq }< \zeta_{\rm L} \le \zeta_{\rm ad}, & \mbox{\ thermal timescale mass transfer. }  \\
  \mbox{if  } \zeta_{\rm L} \le \min(\zeta_{\rm eq},\zeta_{\rm ad}), & \mbox{\ secularly stable mass transfer. } 
\end{array}
\right.
\label{zetaeq}
\end{equation}

\noindent If the new equilibrium radius after the mass loss 
is smaller than the new Roche lobe radius, $\zeta_{\rm L}\le \min(\zeta_{\rm eq},\zeta_{\rm ad})$, 
the donor remains in both thermal and hydrostatic equilibrium all the time.
The mass transfer is then secularly stable and proceeds on a timescale that is independent of the donor envelope's 
thermal and dynamical  responses, but is dictated by other processes (e.g.,  orbital
angular momentum\index{angular momentum} evolution, or nuclear evolution\index{nuclear evolution} of the donor).

If new equilibrium radius is larger than the Roche lobe radius, but $\zeta_{\rm L}\le \zeta_{\rm ad}$, 
the mass transfer will be driven  by continuous thermal readjustment that would keep pushing the star's radius so as to overfill its Roche lobe.
It is a {\it stable} mass transfer although it proceeds from a donor that is out of thermal equilibrium.
Indeed, suppose that the mass transfer is unstable, and has started to increase exponentially. 
Then the reaction of the donor will be determined by $\zeta_{\rm ad}$, and since the mass transfer
is dynamically stable, the donor will shrink inside its Roche lobe, decreasing the mass transfer rate.

\vskip0.2cm

\noindent{\bf Superadiabatic readjustment} Although it has become a norm to analyse the donor's response on two widely
separated timescales, in \emph{real} stars the thermal readjustment of the outer layers of a donor could occur on a very short timescale.
If $\tau_{\rm KH, surf} < \tau_{\rm ML} = M_{\rm d}/ \dot M_{\rm d}$, 
the surface layers are capable to keep rebuilding their thermal structure during the mass transfer \cite{woodsmt11}.
While hydrodynamic readjustment still takes place, this readjustment can not be treated as adiabatic.
The donor's response in this case, $\zeta_{\rm sad}$, is defined by the entropy profile that the donor
is constantly restoring at its surface during the mass loss, and is function of the mass loss rate.
At very high mass rates, $\zeta_{\rm sad} \rightarrow \zeta_{\rm ad}$. At the  mass loss rates closer
to thermal, $\dot M_{\rm KH} = M_{\rm don}/ \tau_{\rm KH}$,  the response $\zeta_{\rm sad} \rightarrow \zeta_{\rm eq}$.
Since donor's response is not anymore adiabatic, the mass transfer then could be stable even if $\zeta_{\rm L} > \zeta_{\rm ad}$.
$\zeta_{\rm sad}$ is function of the mass loss rate and can only be obtained by performing stellar calculations.

\begin{equation}
\zeta_{\rm sad} (\dot M) \equiv \left (\frac{\partial \log R_{\rm d}}{\partial \log M} \right )_{\rm \dot M} 
\Rightarrow \mbox {if } \zeta_{\rm L} \le \zeta_{\rm sad}, \mbox{the mass transfer is dynamically stable}.
\label{zetasad}
\end{equation}

\subsection{Envelope's Structure}

\subsubsection{Adiabatic Response}

\label{sec_ad}

\vskip0.2cm
\noindent{\bf Convective envelope} For a non-degenerate and fully ionised ideal gas, during 
an {\it adiabatic} motion of a convective eddy, pressure and density  
are related as $P \rho^{-\Gamma_1} = {\rm constant}$, with the adiabatic exponent $\Gamma_1=5/3$.  
Since this condition is satisfied everywhere inside a convective zone, a fully convective star 
can be considered as an isentropic polytrope\index{polytrope} of index  $n=1.5$. 
Giant stars\index{red giant}, which possess large outer convective envelopes, are also often described 
as polytropes. Since giants are not fully convective and contain a radiative core, 
additional considerations could include composite polytropes, where the inner part was treated as a polytrope of $n=3$, and
condensed polytropes, while the core is a point mass.

The property of the  polytrope of index $n=1.5$ is that its radius increases when its masses decreases, $\zeta_{\rm n=1.5}=-1/3$. 
This response is the adiabatic response of a 
fully-convective low-mass main sequence star, or a limiting case of the adiabatic response of a giant star.
With hydrodynamics simulations, it was also found that the presence of the core, 
as in the case of either composite or condensed polytropes, increases the value of $\zeta_{\rm ad}$ \cite{Hje87, Sob97},
implying that, in accordance to the criterion (\ref{zetaad}), the stability of the mass transfer {\it increases} 
as the star evolves on the red giant branch. If $M_{\rm c}$ is the core mass of the donor and $m=M_{\rm c}/M_{\rm d}$, then (see \cite{Sob97}):

\begin{equation}
\zeta_{\rm ad} = \frac{2}{3} \left ( \frac {m}{1-m} \right ) - \frac{1}{3} \left ( \frac{1-m}{1+2m}\right ) - 0.03m +0.2 \left [ \frac{m}{ 1+ (1-m)^{-6}} \right ]\ .
\label{zetasob}
\end{equation}

\begin{figure}
\includegraphics[width=59mm]{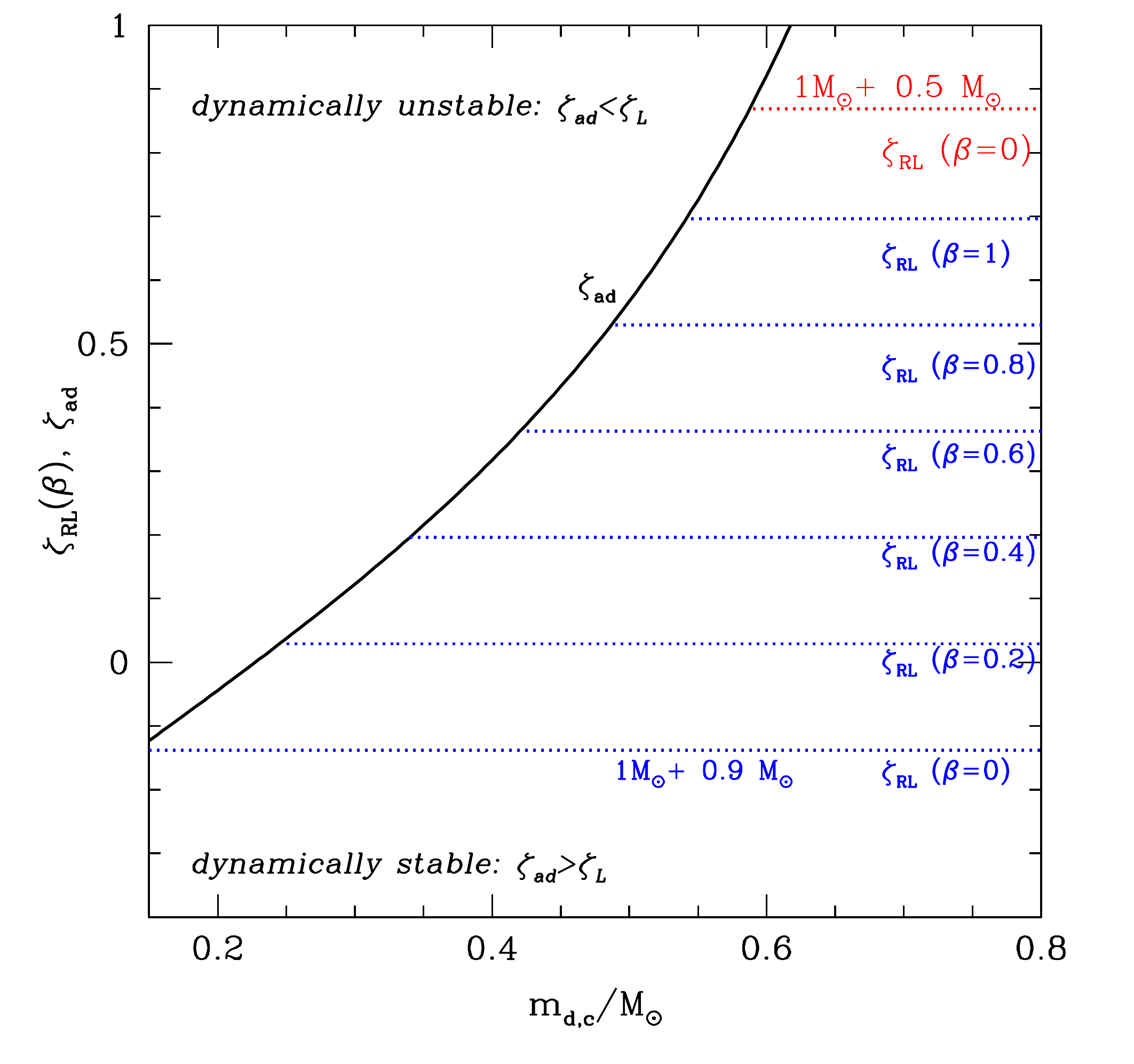}\hskip-0.8cm
\includegraphics[width=59mm]{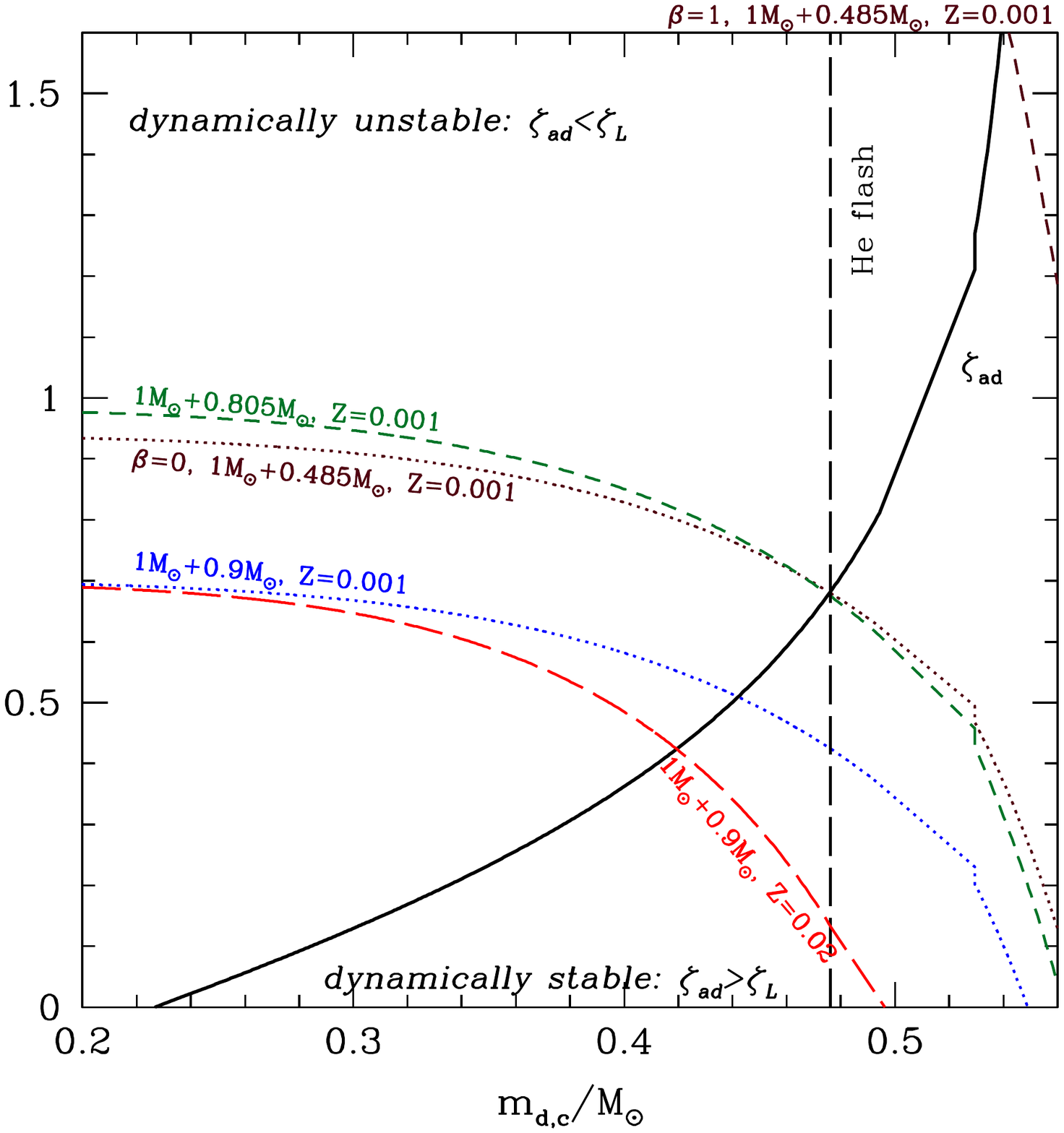} 
\caption{Comparison of adiabatic and Roche-lobe mass-radius exponents\index{mass-radius exponent}. {\bf Left panel}: a red giant of 1 M$_\odot$ of solar metallicity $Z=0.02$ evolved without any mass loss, with companions of 0.9M$_\odot$ and 0.5 M$_\odot$, as a function of the red-giant core mass and for different cases of mass conservation $\beta$. {\bf Right panel}: red giants of 1 M$_\odot$ with $Z=0.02, 0.001$, evolved with wind-mass loss. Dashed line indicates where giants evolve through the He flash\index{helium flash}. $\zeta_{\rm ad}$ is calculated using (\ref{zetasob}).}
\label{zeta_cons}
\end{figure}

\noindent The effect of the core is non-negligible: $\zeta_{\rm ad}$ is positive when $m\ga 0.2$ (see the Fig.~\ref{zeta_cons}).

\vskip0.2cm

\noindent{\bf Radiative envelope} 
By definition, radiative layers are stable against convection,  and, hence, the temperature gradient is 
$\nabla \left ( \equiv \frac{\partial \ln T }{\partial \ln P} \right) < \nabla_{\rm ad}=0.4$.
For an ideal gas, it can be shown that 

\begin{equation}
s(m) \propto \rho^{(\frac{5}{3}\nabla-\frac{2}{3})/(1-\nabla)}.
\label{en_nabla}
\end{equation} 
If the temperature gradient  $\nabla<0.4$, the entropy is growing towards the surface.
If some mass is removed, then a layer with a lower entropy is exposed.
In an adiabatic case, this implies that if a star attains hydrostatic equilibrium and regains
the same surface pressure as before, the density of the outer layers will be larger than previously, 
and the donor shrinks. Hence for radiative donors, $\zeta_{\rm ad} \gg 0$.

\vskip0.2cm

\subsubsection{Equilibrium Response}

The equilibrium response depends on the evolutionary state of the donor.
For low-mass giants, the thermal structure is almost independent from the envelope mass and is a function of the core mass, 
so $\zeta_{\rm eq}=0$.
For main sequence stars, the response is dictated by the mass-radius relation and is usually positive for all donor types. 
E.g., for zero-age main sequence  stars\index{ZAMS} $\zeta_{\rm eq}$ is at least 0.57 \cite{dem91}, and it grows as the donor approaches the terminal main sequence.
Note, that these values are applicable only for the start of the mass transfer.

\subsubsection{Superadiabatic Response}

This is the response of the donor on the mass loss that proceeds on a timescale longer than $\tau_{\rm dyn}$ but shorter than $\tau_{\rm KH}$.
Arguably, this is the most important response for determining the mass transfer stability. 
In fact, in stellar codes, the decision on whether a star is experiencing dynamically unstable mass loss
is often done when the mass loss is still $\la 10^{-3}$~M$_\odot$ per year, but only because stellar codes often 
can not converge on a star that looses mass at a faster rate. 
Note that this mass transfer rate is still several orders of magnitude slower than if it would be 
on a real dynamical timescale, which is presumed for an ideal adiabatic response.

The understanding of superadiabatic response is most important for the cases of the mass transfer 
which are deemed to be unstable if adiabatic response is considered.
Radiative donors have $\zeta_{\rm ad} \gg 0$ and are widely accepted to be more stable than convective donors,
so we will consider here the superadiabatic response for convective donors only.

Why it is called ``superadiabatic response''?
The entropy profile in convective stars is not as simplistic as was discussed above in \S\ref{sec_ad}.
Closer to the surface, the energy transport by convection is becoming not efficient enough to transport all the luminosity, 
 and the transfer of energy by radiation starts to play a role (the region is still convectively unstable). 
The temperature gradient is now determined by both energy transport
mechanisms, and its value is now somewhere between the adiabatic temperature gradient and 
the radiative temperature gradient $\nabla_{\rm ad} < \nabla < \nabla_{\rm rad}$, hence, it is called \emph{superadiabatic}\index{superadiabatic}.
In this region, the entropy of the material is becoming {\it smaller} than in the  
adiabatic convective envelope\index{convective envelope} beneath it --- see also (\ref{en_nabla}). 
The convective envelope appears to be covered by a thin superadiabatic blanket.

\begin{figure}
\includegraphics[width=119mm]{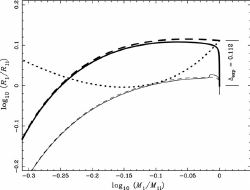}
\caption{Adiabatic mass-loss sequences for \emph{true} (thick solid lines) stellar models and \emph{pseudo-models} (i.e. adiabatic  models; thick dashed lines) of a 1 M$_\odot$ star near the tip of the giant branch. $\nabla_{\rm exp}$ is the prompt initial expansion resulting from the removal of superadiabatic surface layers. Also shown are the Roche lobe radii\index{Roche lobe radius} within true and pseudo-models (thin solid and dashed lines, respectively) at which the mass-loss rate would reach $\dot{M}_{\rm KH}$.  
Critical initial mass ratio at which mass transfer rate reaches $\dot{M}_{\rm KH}$ for the pseudo-model is shown by the heavy dotted curve. 
(reproduced from \cite{Ge2010} by permission of the AAS). }
\label{zeta_ad_ge}
\end{figure}

If mass is removed from this entropy profile in an adiabatic regime, the envelope is momentarily expanding
by a large fraction of its radius. For example, a 1 M$_\odot$ giant would expand by as much as $\sim25\%$  (see Fig.~\ref{zeta_ad_ge}).
This dramatic expansion after removing a tiny mass from the surface ($<10^{-4}$~M$_\odot$) demonstrates that
a simplified star with a constant entropy profile in the envelope, up to the surface, 
as was considered in \S\ref{sec_ad}, finds a very different hydrostatic equilibrium than a normal giant of 
almost the same mass, and the role of the superadiabatic layer in finding that new equilibrium, 
and hence $\zeta_{\rm ad}$, could be enormous.
The applicability of the adiabatic approach is therefore quite questionable.

Woods \& Ivanova \cite{woodsmt11} investigated how giant envelopes are responding in realistic stellar models 
when fast mass loss proceeds. 
They found that if the superadiabatic layer is resolved, the mass transfer could be stable 
in systems with larger mass ratio than would be predicted by the polytropic approach (in the considered systems, $q_{\rm crit}\sim 1.5$ instead
of $\sim 0.8$ as derived when using the adiabatic approximation).
They also found that real giants will often contract while the adiabatic response, given by (\ref{zetasob}), predicts an expansion. 
A giant would start to expand  only at the very high mass loss rate, $\ga 0.1$~M$_\odot yr^{-1}$, and this expansion is much smaller 
than predicted by the adiabtic approach --- less than $1\%$ of its radius.

These effects were observed in two fairly different stellar codes.
One code is Lagrangian, a standard \emph{Henyey-type} code\index{Henyey code} developed by Podsiadlowski \& Rappaport \cite{podsi02} from 
what was originally described in Kippenhahn, Weigert \& Hofmeister \cite{Kip67}; the current state of the code and input physics can be found in \cite{Iva03,Iva04}.

The second code is a non-Lagrangian code  {\it STARS/ev}\index{STARS code}, originally developed by Eggleton 
(\cite{1971MNRAS.151..351E,1972MNRAS.156..361E,1973MNRAS.163..279E,1973AA....23..325E}, 
and with updates as described in \cite{1995MNRAS.274..964P,2008AA...488.1007G,2011ascl.soft07008E} and references therein). 
The contraction of the giants upon the mass loss was also confirmed by Passy, Herwig \& Paxton \cite{2012ApJ...760...90P} 
using yet another stellar code, {\it MESA}\index{MESA code} \cite{MESA1, MESA2}. 

Nonetheless, even though three different stellar codes reproduced qualitatively the same result -- 
the giant's contraction upon rapid mass loss -- 
it is not clear if {\it any} of the existing stellar codes is capable of producing a valid result in this regime.
The caveat is that the continuous  re-establishment of the surface 
superadiabatic layer proceeds on a timescale shorter than a convective eddy turnover time.
In this case  the commonly accepted convection treatment (the mixing length theory) 
may not be further applicable  \cite{pavl_casca}.

\subsubsection{Sudden Change of the Donor's Response -- Delayed Dynamical Instability} \index{delayed dynamical instability}

This is the case of a dynamical instability\index{dynamical instability} that  follows  a period of a stable mass transfer.
It occurs if the donor's response $\zeta_*$ {\it suddenly} drops during the mass transfer and become smaller than $\zeta_{\rm L}$.
It usually takes place in initially radiative donors, when the mass loss ripped enough of the donor's material to reach 
layers where the convective core was and where the entropy is still nearly constant 
(for a more extended discussion, see e.g. \cite{Ge2010}). 

Since the first phase of the mass transfer proceeds roughly on a thermal timescale and its rate is a function of the mass ratio, 
the entropy of the inner layers might change compared to the initial profile. 
Hence the response $\zeta_*$, as a function of the current donor mass,  will depend on the history of the mass transfer.
The prediction of when exactly the delayed dynamical instability (DDI) could start is impossible by considering the initial donor only, but
requires detailed stellar modeling.

Some rough values of $q_{\rm crit}$ separating the binaries that will experience a DDI and those proceeding with a stable mass transfer
are known from previously performed calculations (for example, see \cite{Iva04, Ge2010})
but vary from code to code; $q_{\rm crit} \ga 3$.

\label{sec_ddi}

\subsubsection{Donor's Pulsations}

Red giants\index{red giant} are known to have intrinsic pulsations\index{pulsation} (for a review, see \cite{Dalsgaard2011}, and the references therein).
When hydrostatic equilibrium is perturbed due to mass loss, and a giant attempts to restore it, these intrinsic  pulsations might be amplified.
Indeed,  it was shown \cite{2013IAUS..290..293P} that the initiation of the mass loss causes mass-loss induced pulsations, 
where the radius amplitude could be about $1\%$ of the star's radius, $\delta R\sim 0.01 R$.  
While this radius increase may seem small, the mass loss rate during RLOF is a strong function of the fractional radius excess of the donor, $\dot M \propto (\delta R/R)^3$. Hence, mass-loss induced pulsations could affect the rate of mass transfer during RLOF as well as the initial stability of the mass transfer.

\section{The Donor's Response and the Consequences for the Mass Transfer Stability}

\subsection{Initial Stability}

\subsubsection{Fully Conservative Mass Transfer}

In a binary star, the initial mass ratio $q>1$, implying that the initial 
$\zeta_{\rm L}>0.46$ in the case of a fully conservative mass transfer (see (\ref{zetal_cons}) above). A comparison with values of $\zeta_{\rm ad}$ shows that donors with convective envelopes are expected to start the mass transfer in a dynamically unstable way. The dynamical instability could be avoided in a second episode of the mass transfer, if $q<0.8$ at the second RLOF.

Donors with radiative envelopes have $\zeta_{\rm ad}\gg 0$ and they can avoid dynamical instability at the start of mass transfer for a large range of mass ratios\index{mass ratio}. 
Mass transfer will unavoidably be dynamically unstable from the very beginning for systems with radiative donors and
$q_{\rm crit}\ga10$ due to the Darwin instability.

\subsubsection{Non-Conservative Mass Transfer}

As follows from (\ref{nonc_eq1}) and (\ref{nonc_eq2}), if mass is lost from the system, the stability of the mass transfer is increasing. This can be demonstrated using a simple ``toy'' model, where $\zeta_{\rm ad}$ (obtained using \ref{zetasob}) is compared  to $\zeta_{\rm L}$, for a 1 M$_\odot$ giant with two different companions, 0.9 M$_\odot$ and 0.5 M$_\odot$, and assuming that different fractions of transferred mass can be accreted (see Fig.~\ref{zeta_cons}, left panel).  It can be seen that a low-mass giant in a binary with a 0.9 M$_\odot$ companion cannot have both stable and fully conservative mass transfer while it is a red giant (core mass $M_{\rm c}\la 0.47$~M$_\odot$). The stability of the mass transfer can only be achieved if $\beta \la 0.6$. 
A fully non-conservative mass transfer ($\beta=0$) will be stable in a binary consisting of a  1 M$_\odot$ + 0.9 M$_\odot$ starts, whatever the 
red-giant core mass. The critical mass ratio for $\beta=0$ is larger by $\sim 40\%$ than for a fully conservative case for the same donor.

\subsubsection{Mass Loss Prior to RLOF}
The ``toy'' model used a star evolved without wind mass loss. Loosing mass through winds prior to RLOF reduces the relative envelope mass, increases the donor's $\zeta_{\rm ad}$ and makes the mass transfer stable for a larger range of donor's radii (or core masses\index{core mass}, which are tightly related to the radii for low-mass giants\index{stellar radius}). For example, a red giant of initial 1 M$_\odot$ could have a stable and fully conservative mass transfer onto a donor of 0.9 M$_\odot$ once it developed a core $M_{\rm c}\ge 0.42$~M$_\odot$  (see Fig.~\ref{zeta_cons}, right panel). 
Low-mass giants of lower metallicity, as in metal-poor globular clusters, would loose less mass through wind. Hence, the stability of the mass transfer in metal-poor globular clusters can only be achieved for a smaller range of initial binary parameters.

\subsection{Stability of the Ensuing Mass Transfer}

\begin{figure}
\includegraphics[width=119mm]{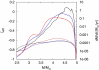} 
\caption{Evolution of the effective radial response of the donor (solid lines) and mass loss $\dot{M}$ (dashed lines) during the mass transfer, 
for a 5 M$_\odot$ + 5 M$_\odot$ binary with a donor radius of 50 R$_\odot$ (black, highest peak), 68 R$_\odot$ (blue, middle), and 85 R$_\odot$ 
(red, lower) with a point mass companion. (reproduced from \cite{woodsmt11} by permission from the AAS)}
\label{zeta_giant_time}
\end{figure}

In order to have a dynamically stable mass transfer, 
the condition for stability must be satisfied not only
at the start of the mass transfer, but 
during the whole duration of the mass transfer. 
In population synthesis studies, especially where a parameterisation 
is used for stellar codes to speed up calculations,
the instability is evaluated only at the start of the mass transfer 
(e.g. \cite{iva2002MNRAS.329..897H,2008ApJS..174..223B}). It is partially justified, as 
in simple cases $\zeta_{\rm L}$ is decreasing as the mass transfer proceeds 
(e.g., if the mass transfer is fully conservative), 
hence the stability of the mass transfer is expected to only increase after it started.

The noticeable exception that is usually taken into account is the DDI.
For DDI, population studies accept some value for $q_{\rm crit}$, and for all systems with $q>q_{\rm crit}$
the mass transfer is considered to be unstable (e.g., \cite{2008ApJS..174..223B})
(note, however, that the mass transfer in reality would become unstable only after a significant fraction of the donor is transferred on its thermal time scale).
While population synthesis codes accept some strict value, 
the exact value of the critical mass ratio varies between the code and depends on considered systems. 
For example,  Ivanova \& Taam \cite{Iva04} found $q_{\rm crit}\approx 2.9-3.1$ on one sample of donors, 
while Ge et al. \cite{Ge2010} found this value to be higher, $q_{\rm crit}> 3.5$, 
on another sample of donors\footnote{we note also that in their approach DDI was based on a frozen entropy profile}.
This classic DDI scenario is assumed to happen to donors with radiative envelopes and convective cores.

The change of the donor's response, however, could occur in other cases, not only in the classic DDI case.
In convective giants, $\zeta_*$ is not constant during the mass transfer (see Fig.~\ref{zeta_giant_time}),
and its value can go down after an insignificant fraction of the envelope has been already  transferred. 
The particular example shown in Fig.~\ref{zeta_giant_time} adopted  fully conservative mass transfer but was nonetheless stable,
despite the fact that the polytropic argument would predict all of them being unstable --- and any non-conservation would improve stability even further.
However, the important point here is that even in giants\index{red giant} with convective envelope\index{convective envelope} the response of their envelope 
is changing with time.
Subsequently, it is possible that for any convective donor there is a $q_{\rm crit, DDI}$ such that for mass ratios $q>q_{\rm crit, DDI}$ 
but $q<q_{\rm crit}$, the DDI arises. In these systems, a period of initial stable mass transfer will 
increase the survivability of the binary once dynamical instability (common envelope) occurs, since a fraction of the envelope
was removed at the expense of ``nuclear'' energy during the donor's evolution.

\subsection{Stable or Not Stable?}

The determination of whether dynamical instability takes place  at the start of the mass transfer
or a DDI takes place during the mass transfer, hits the same problem: even 
detailed binary stellar codes are numerically limited by their abilities to converge a donor that experiences
fast mass loss. The mass transfer is usually declared to be dynamically unstable if its value exceeds some 
adopted critical value for the mass transfer rate, different for various codes, and depends on what a code is capable of handling.
Taking a too low value may lead to wrong conclusions -- e.g., $10^{-3}$~M$_\odot yr^{-1}$. Note that this
is a mass transfer rate that is by several orders of magnitude larger than in the dynamical regime. This
inability of a stellar code to converge does not imply that dynamical mass transfer --- in other words, that a star is not capable anymore to remain in hydrostatic equilibrium --- has indeed occurred. There are mass transfer sequences where thermal time scale mass transfer rate from a giant donor 
approaches $10^{-2}$~M$_\odot yr^{-1}$ (still 100 times slower than dynamical timescale), 
and then the donor recovers and continues the mass transfer on a nuclear time scale. 

In addition to problem with the convergence of the donor, the converged solution might be intrinsically incorrect, depending on what physics was taken into account. While many stellar codes now include a component that takes into account structural changes on a dynamical timescale, 
convection is for instance considered in the approximation known as
mixing length theory\index{mixing length theory}, which can break down when significant changes happen on a timescale comparable to a convective eddy turnover time \cite{pavl_casca}.

\subsection{Three-dimensional Problem}

The RLOF process in a binary system is intrinsically a three-dimensional (3D) problem, while all the possible criteria of mass transfer 
instability described above are valid only for one-dimensional, spherically symmetric stars.
While there are simulations of three-dimensional behaviour of the stream after it had already left $L_1$ (e.g. \cite{1997MNRAS.292..321Y}), 
there is no fully self-consistent simulations yet of the start of the RLOF in 3D, where the RLOF is truly driven by binary evolution.

Most 3D studies of RLOF binaries are rather devoted to the understanding of the common envelope evolution. Hence, the topic
under investigation is not the stability, but instability and its consequences.
Since evolution-driven RLOF takes place on a timescale much longer than the dynamical timescale (hydrodynamical codes are not 
capable of treating events on timescale much longer than a dynamical timescale unless proper stellar physics, especially
for energy transfer, is added),
the start of the common envelope is usually artificially enforced. Example of enforcement include  
starting with initial Roche lobe overflow, or starting with initially asynchronised companions 
and allowing tidal forces to drive the coalescence
\cite{2012ApJ...744...52P,2012ApJ...746...74R}.

Recently, one of the Smoothed Particle Hydrodynamics (SPH)\index{Smoothed Particle Hydrodynamics} codes was specifically modified to study stability of the mass transfer, by
introducing a proper treatment of synchronised binaries, including cases where a companion could be just reaching its 
RLOF  \cite{2011ApJ...737...49L}.
In recent calculations using that code Avendano Nandez et al. (in preparation) considered a binary system that consisted 
of a 8 M$_\odot$ giant and 5 M$_\odot$ black hole.
The expectation for such a system (the mass ratio $q=1.6$ is almost twice the critical mass ratio for a donor with convective envelope) from 1D stability 
analysis is very firm --- this system is destined to start the common envelope\index{common-envelope evolution} phase, i.e. a dynamical unstable RLOF. 
However, in the 3D simulations, this binary failed to start the dynamically unstable mass transfer.
In six years (physical time) of RLOF,  the average mass transfer rate in the simulation was $\sim 0.02$~M$_\odot yr^{-1}$; 
most of the transferred mass did not stay within the black hole's Roche lobe: 0.084 M$_\odot$ was ejected from the binary to infinity and
0.025 M$_\odot$ went into a circumbinary disc. This example clearly suggests that criteria based on 1D stellar calculations cannot be final, and mass transfer in real 3D stars could be stable for a much larger range of mass ratios, whether this is due to
strong liberal evolution\index{liberal evolution}, or different from spherically symmetric case donor's response.

\section{The Accretor's Response and Consequences for Mass Transfer Stability}

During RLOF, the donor's material, once leaving the nozzle at the $L_1$, proceeds towards the accretor in the form of a stream. The 
stream's trajectory is more-or-less ballistic (for vertical structure and trajectory of an isothermal stream see \cite{1976ApJ...207L..53L} 
and for an adiabatic stream see \cite{ips02}).
The material in the stream: 
\begin{enumerate}
\item has the chemical composition of the donor;
\item carries angular momentum from $L_1$; 
\item {\it may} have the entropy of the donor's material.
\end{enumerate}
The transfer of a donor's composition\index{chemical composition} may create a specific signature that could help to identify a star formed through mass transfer\index{mass transfer}, but it does not affect the stability of the ongoing mass transfer. The stream's angular momentum\index{angular momentum} and  entropy\index{entropy} may, however, affect it.

\subsection{The Stream's Angular Momentum}

After leaving $L_1$, the stream conserves its angular momentum  $J_{\rm s}$ for as long as its trajectory is
 unperturbed from interacting with some other material.
Depending on the sizes of the accretor and of its Roche lobe, the stream could either  collide with
itself, forming an accretion disc\index{accretion disc} around the accretor at the circularisation radius\index{circularisation radius} 
$R_{\rm circ} = J_{\rm s}^2/ GM_{\rm a}$, or, if the accretor's radius $R_{\rm a } > R_{\rm circ}$,  the stream
can hit the accretor directly, transferring to the accretor its angular momentum.

If an accretion disc is formed, as can be expected in relatively wide binary systems that proceed with a case B or a case C mass transfer or if the accretor is a white dwarf\index{white dwarf}, a neutron star\index{neutron star}, or a black hole\index{black hole}, 
the accretor is not necessarily gaining angular momentum.
But in case of a direct impact, the accretor can be quickly spun up.
It can be estimated  that the accretor can reach \emph{critical rotation}\index{critical rotation} --- defined when the centrifugal force equals the gravitational force and the star is no more bound --- after accreting just 
a few per cent of its mass \cite{1981A&A...102...17P}.
This raises for blue stragglers formed via case A or early case B mass transfer
the same problem as in the case of blue stragglers formed via collisions: how to get rid of the
angular momentum? (not to mention that the physics of stars that are close to critical rotation is
poorly understood, see, e.g., for a review 
\cite{2008IAUS..250..167L}).

A possible solution is that the stream is not able to transfer all its angular momentum to the accretor.
Indirect evidence, for instance, suggests that it is hard to recreate the distribution of observed Algol systems with mass ratios in the range $0.4\div0.6$ with fully conservative mass transfer   \cite{2008AA...487.1129V}. Also,  magneto-hydrodynamic simulations presented in
\cite{2012MNRAS.427.1702R} showed that the stream that would be normally expected to have a direct impact, is deflected, where the deflected formation is taking away most of the stream angular momentum. In addition to resolving the problem with the excess of the angular momentum in the accretor, stream's deflection also stabilizes the mass transfer.

\subsection{The Accretor's Response}

Let us consider the case of stable mass transfer, operating on a thermal timescale.
For as long as the mass of the donor remains larger than the mass of the accretor, 
the accretor's thermal timescale is longer than that of the donor (this is valid at the start of both cases A and B mass transfers,
or for as long as both companions remain main sequence stars\index{main sequence star}).
The accretor, just like the donor, has to readjust to its new mass,
but the timescale at which it has to readjust is shorter than its thermal timescale.
The accretor, hence, will be brought out of its thermal equilibrium.

This accretor's response on a timescale shorter than its thermal timescale 
can be considered as the reverse of a rapid mass loss from the donor (adiabatic response,
though in reality this response will be closer to superadiabatic response):
 if the accretor has a radiative envelope, it can be expected to expand upon the mass accretion,
while if the accretor has a convective envelope, it can be expected to contract upon the mass accretion.
For example,  an initially radiative donor is expected to be both more luminous\index{overluminous star} and more expanded 
than a star of the same mass in thermal equilibrium.
For sufficiently high accretion rates, an accretor then might even overfill its Roche lobe\index{Roche lobe}, forming a contact binary\index{contact binary}.

A contact binary could also be formed during nuclear timescale mass transfer, when 
then accretor's expansion takes place due to its evolution off the main sequence.
Other possibilities, as was mentioned before, include a white dwarf's reincarnation into a red giant
and the formation of a ``trapping'' envelope around a neutron star.
However, none of these situations leads to formation of a blue  straggler.

\subsection{Donor's Entropy and the Accretor's Response}

The accretor's response onto mass accretion, as described above, has been confirmed by binary calculations (e.g., 
\cite{2001ApJ...552..664N}). These calculations, however, 
made two simplifications: the accreting material had
the surface composition and entropy (as well as temperature) of the surface material of the accretor, not of the donor. 
More recent evolutionary codes include \emph{thermohaline mixing}\index{thermohaline mixing} (e.g. \cite{2008MNRAS.389.1828S}), 
and are capable to add material of different chemical composition in a proper way.
Adding material with different entropy remains a problem. It is also unclear what exactly 
should be the entropy of the accreting material: 
a) the stream does not necessarily keep the same entropy as the donor  while it is in the ballistic phase;
b) the stream could be shock-heated during its impact with the accretor.

So why is the entropy of the material during the accretion  
(and hence the entropy profile of the accretor) a problem? 
It must be clarified that, especially in the case of blue straggler formation, 
this is an entirely different issue from the entropy profile of a collision product.  
In the latter case,  the entropy profile of the collision product is created 
by \emph{simplified entropy sorting} \cite{2002ApJ...568..939L}, 
with mandatory positive entropy gradient. 
This profile then defines an initial, out-of-thermal equilibrium, 
stellar model of a blue straggler, 
which is then further evolved with a stellar code.

On the other hand, accretion of material with different entropy creates 
an entropy discontinuity in the accretor's envelope. The
physics of this discontinuity is closely related to the 
problem of a transitional layer in contact binaries and is
not yet fully understood  (see, e.g., discussion in \cite{1980ApJ...239..937S}).

\section{How Well Do We Understand Stable Mass Transfer?}

\label{sec_howwell}

\begin{figure}
\includegraphics[width=119mm]{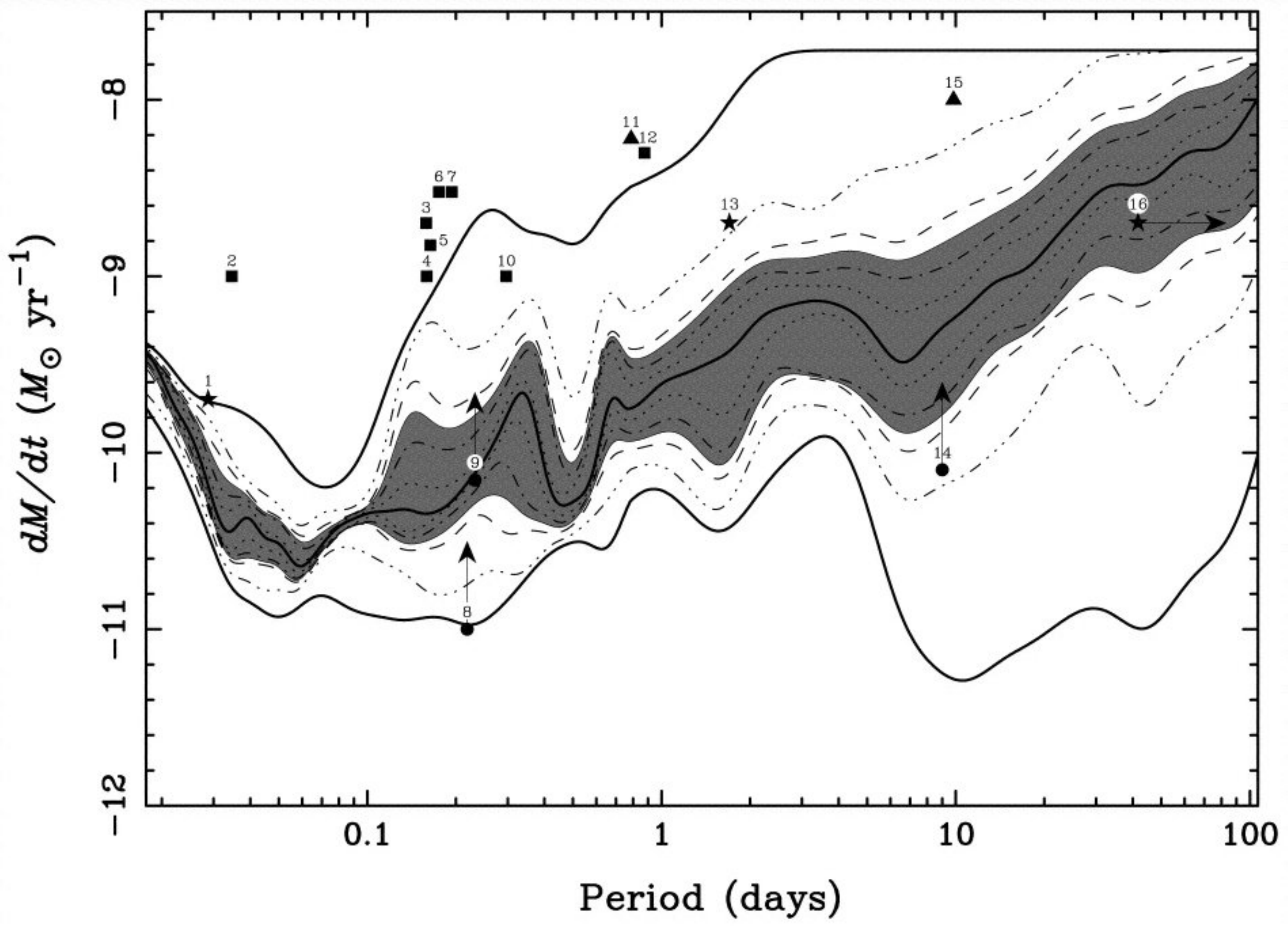} 
\caption{Cumulative (smoothed) probability distribution for the mass accretion rate onto a neutron star primary as a function of orbital period for 100 binary calculations (with equal weighting for all sequences). The thick, solid central curve gives the median mass accretion rate; the pairs of curves moving progressively outward from the median curve include 20\%, 40\%, 60\%, 80\%, and 98\% of the distribution, respectively. The shaded region contains 50\% of all systems around the median. The symbols indicate the mass transfer rates of selected observed X-ray binaries (triangles: Z sources; squares: atoll sources; stars: X-ray pulsars; circles: systems with accretion disc coronae [lower limits]). The individual systems are (in ascending order of orbital period, as given in parentheses) (1) 1626-67 (0.69 hr), (2) 1916-053 (0.83 hr), (3) 1636-536 (3.8 hr), (4) 0748-676 (3.82 hr), (5) 1254-690 (3.93 hr), (6) GX 9+9 (4.2 hr), (7) 1735-555 (4.65 hr), (8) 2129+470 (5.24 hr), (9) 1822-37 (5.57 hr), (10) 1658-29 (7.11 hr), (11) Sco X-1\index{Sco X-1} (18.9 hr), (12), 1624-590 (21 hr), (13) Her X-1\index{Her X-1} (40.8 hr), (14) 0921-630 (216 hr), (15) Cyg X-2\index{Cyg X-2} (236 hr), and (16) GX 1+4 (greater than 1000 hr). 
(reproduced from \cite{podsi02} by permission from the AAS).}
\label{podsilmxb}
\end{figure}

We can express the change of the donor's radius as due to the evolutionary change and to the thermal equilibrium's response on the mass loss:

\begin{equation}
\frac{\dot R_{\rm d}}{R_{\rm d}} = \left ( \frac{\dot R_{\rm d}}{R_{\rm d}}\right)_{\rm ev} + \zeta_{\rm eq} \frac{\dot M_{\rm d}}{M_{\rm d}}\ .
\end{equation}

\noindent The change of the Roche lobe\index{Roche lobe} radius can be expressed as due to the response on the mass transfer, and to the losses of the orbital angular momentum without the mass transfer:

\begin{equation}
\frac{\dot R_{\rm L}}{R_{\rm L}} = \left ( \frac{\dot R_{\rm d}}{R_{\rm d}}\right)_{\rm aml} + \zeta_{\rm L} \frac{\dot M_{\rm d}}{M_{\rm d}}\ .
\end{equation}

\noindent In the absence of mass transfer, at constant $q$, $R_{\rm L}$ is simply proportional to $a$, and the change of the orbital angular momentum $J$ is $2\dot J / J = \dot a / a$. Then the mass transfer rate can be found from 

\begin{equation}
 - \frac{\dot M_{\rm d}}{M_{\rm d}} = \frac{1}{\zeta_{\rm eq} -  \zeta_{\rm L}} \left [ \left ( \frac{\dot R_{\rm d}}{R_{\rm d}}\right)_{\rm ev} - 
2\frac{\dot J}{J} 
\right ]  \ .
\end{equation}

The best systems to test our understanding of stable mass transfer are low-mass X-ray binaries\index{low-mass X-ray binary} with neutron star\index{neutron star} accretors, where we can obtain the mass transfer rates  from the observed X-ray luminosities with uncertainties to only within a factor of 3 \cite{2007AA...465..953I}. Nonetheless, for low-mass X-ray binaries with non-degenerate donors, the observed mass transfer rates do disagree with theoretically obtained ``most likely'' mass transfer rates by more than an order of magnitude \cite{podsi02}, with observed systems having usually much higher mass transfer rate (see Fig.~\ref{podsilmxb}).

The situation is relatively better in the case of ultra-compact X-ray binaries\index{ultra-compact X-ray binary} -- low-mass X-ray binaries with white-dwarf donors. There, the orbital angular momentum loss\index{angular momentum loss} is via gravitational wave radiation\index{gravitational wave radiation} and is understood much better than angular momentum loss with non-degenerate donors in wide systems. The adiabatic response is simply $\zeta_{\rm ad}=-1/3$. A further improvement was obtained by considering the effect of the final entropy on $\zeta_{\rm ad}$ 
in white dwarf\index{white dwarf} remnants \cite{2003ApJ...598.1217D}. While most of the ultra-compact X-ray binaries do match very well the theoretically obtained mass transfer rates, there are three systems that have mass transfer rate almost two orders of magnitude higher than theoretically predicted, and one system where the mass transfer is much lower than theoretically predicted \cite{heinke2013}.  

Discrepancies between theory and too high mass transfer rates for some ultra-compact X-ray binaries may indicate that the donors are not remnants of white dwarfs \cite{2013IAUS..291..468P,heinke2013}. Otherwise, it can be stated that the theory of the stable mass transfer in systems with a well known mechanism of angular momentum loss (for instance, via gravitational wave radiation) and a well understood simple donor's response (as with degenerate donors where $\zeta_{\rm ad}=-1/3$) agrees with observations very well. However, for the cases applicable to blue straggler formation, where the mass transfer proceeds from a non-degenerate donor, especially from main sequence and early giant donors, the theory of the stable mass transfer is yet far from matching the observations well.

\section{RLOF and  Blue Stragglers Formation}

In this section we discuss {\it how} blue stragglers can be 
formed as a result of RLOF, while the overview of population studies and rates
can be found in Chap. 12.

\subsection{Case A and Early Case B}

Here we consider the cases when the mass transfer resulting in 
a blue straggler formation occurred before the donor developed a deep convective envelope.
From the analysis above, binary systems with low-mass primaries 
are expected to either have thermal timescale mass transfer, or experience DDI\index{delayed dynamical instability}.
A more precise mapping of the actual mass transfer with the 
initial binary parameters (primary mass, mass ratio and orbital period) 
can be done only using binary stellar codes.

Nelson \& Eggleton \cite{2001ApJ...552..664N} performed an extensive study of the Case A\index{Cases A, B, C of mass transfer} 
and early case B mass transfers 
for a wide variety of initial binary systems, 
with primary masses from 0.9 M$_\odot$ to 50 M$_\odot$, 
initial mass ratios up to $\log_{10}q = 0.5 ( q \sim 3.16)$,
and with periods ranging from the contact period at the zero-age main sequence, 
$P_{\rm zams}$ up to $\log_{10} P/P_{\rm zams} = 0.75 ( P\sim 5.6~P_{\rm zams})$. 
This extensive library of models was created to explain Algol systems. 
However, the first episode of mass transfer,
whether it was found to be stable or unstable, creates a blue straggler. 
This study therefore
provides good  insight on when the formed blue stragglers 
remain in a binary, and when they are formed as single blue stragglers.

Analysing this study, it can be accessed that  single 
blue stragglers that can be present  
in globular clusters (with relatively low masses of initial primaries, $<2$~M$_\odot$), 
could be formed both via case A and case B mass transfers.
The first condition that determines whether a binary would merge 
and produce a single blue straggler 
is the initial period $P~\ga~2-3~P_{\rm zams}$, where the larger period 
is for the more massive primaries.
Most of those binaries were found to have rapidly rising above 
thermal timescale case B mass transfer.
The second condition is the initial mass ratio.
First of all, this study adopted the fact that any initial mass ratio that 
is larger than $\sim 3.16$ 
would likely lead to DDI and the merging of the system; these binaries were not calculated.
In the binary calculations done, binaries with primaries 
$\sim 1$~M$_\odot $ and with as small mass ratio as two  
were found to start dynamic timescale case A mass transfer and quickly 
enter into contact, before the 
secondary could accrete enough to appear as a blue struggler. 
At the contact, the calculations were stopped 
and further fate of the systems (merger or long-term contact) was not determined. 

It must be noted that even though the AD and BD cases described above were declared to be 
dynamical, there are not really such.
The adopted criterion for defining  whether the mass transfer is  
dynamical was very soft, 
with the mass transfer rate being just 10 times more than that of 
a thermal timescale mass transfer, 
which for main sequence donors implies many orders of magnitude less than 
realistic dynamical timescale mass transfer.
All the mentioned studies on DDI adopted fully conservative mass transfer, 
and as discussed previously, 
any non-conservation could change the boundary between the formation of 
a blue straggler in a binary and a single 
blue straggler to a lower value of $q$.

\begin{figure}
\includegraphics[width=55mm]{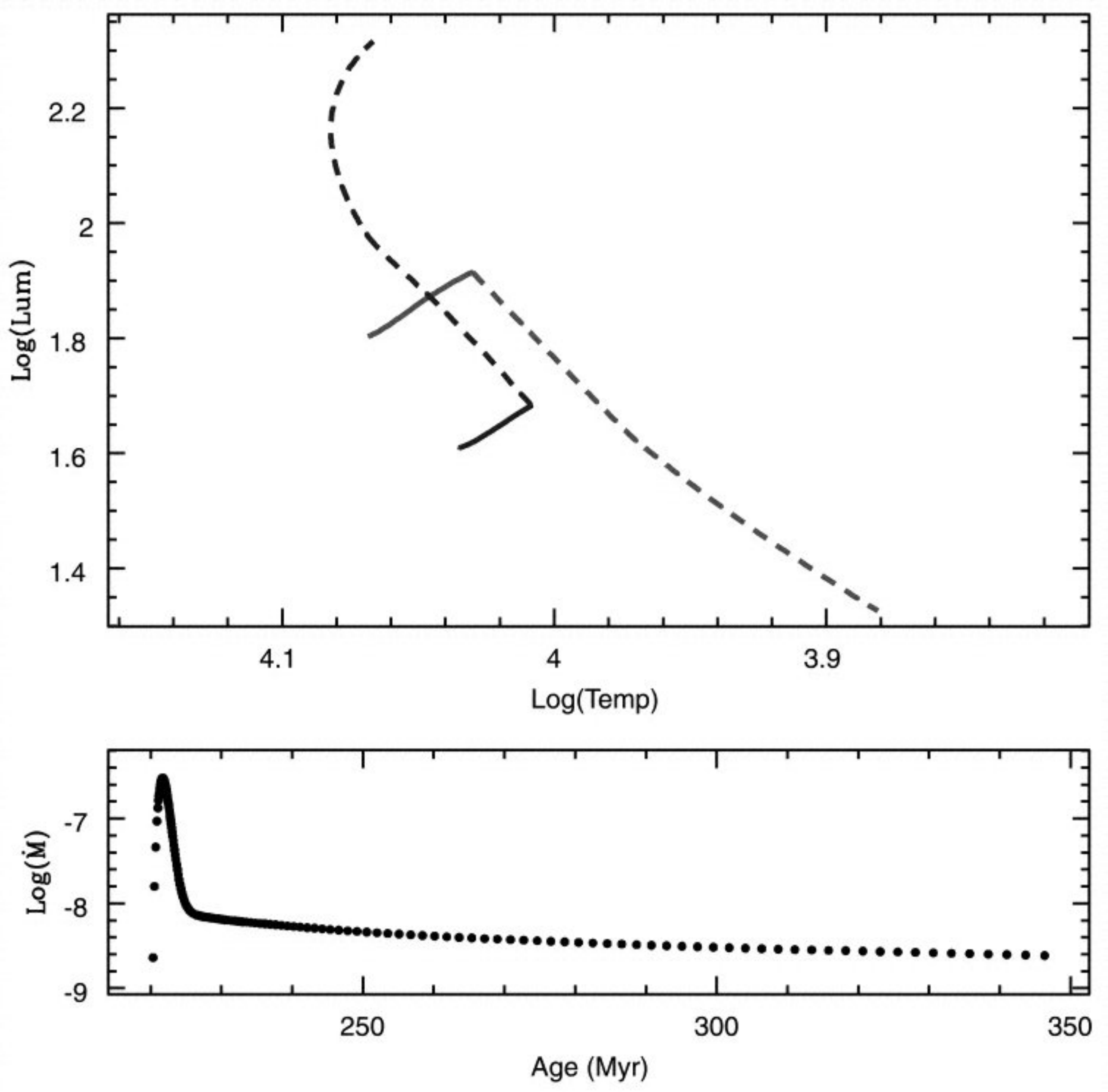}
\includegraphics[width=55mm]{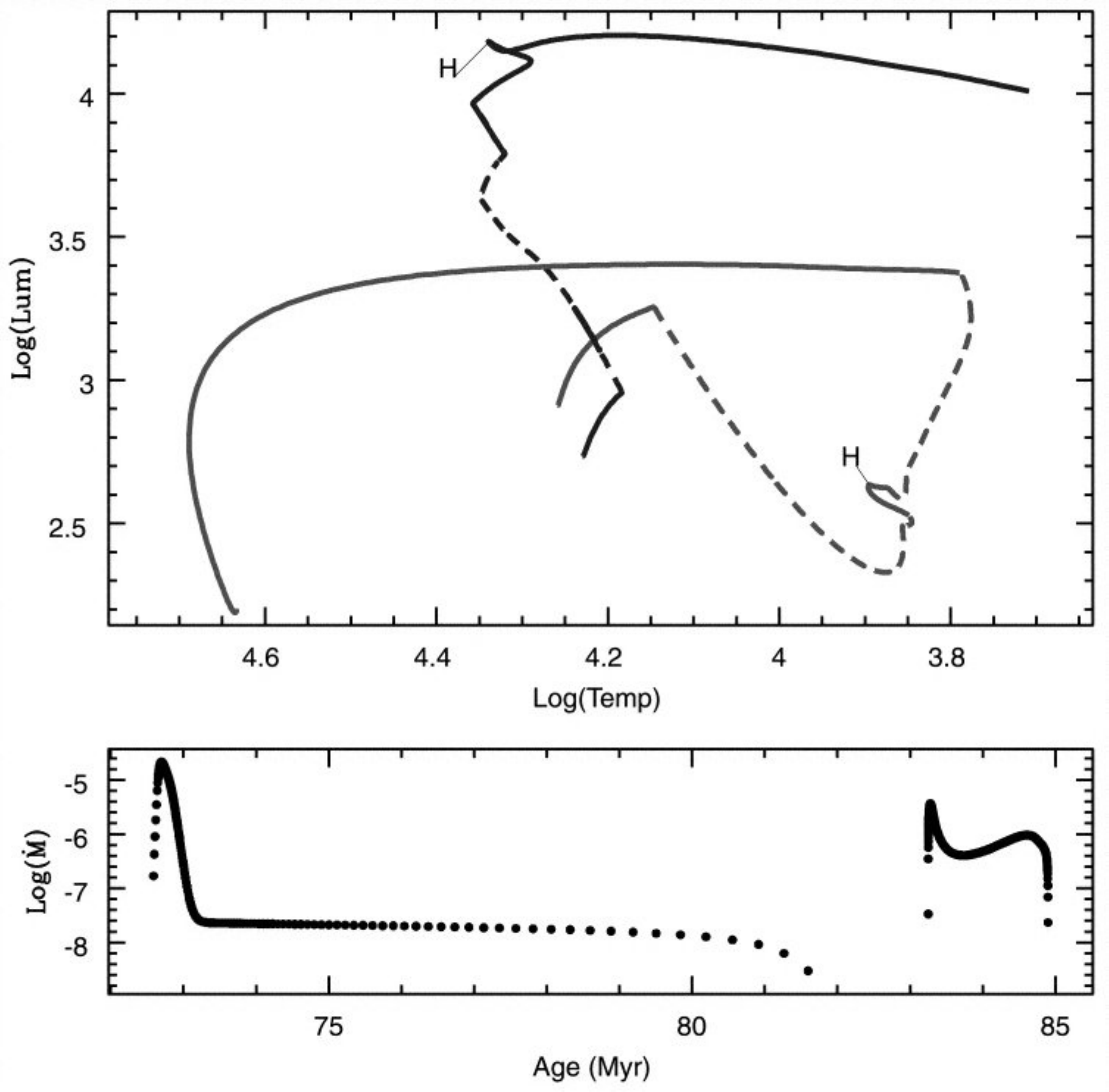} 
\caption{Examples of Case A evolution that result in the formation 
of a contact binary containing a blue straggler.
{\it Left panel:} Case AS --- slow evolution to contact. 
This particular system consisted initially of 2.8 M$_\odot$ + 2.5 M$_\odot$ stars with initial period $\sim 1.6~P_{\rm zams}$. {\it Right panel:} 
Case AL  --- late overtaking.
This particular system consisted initially of 5.6 M$_\odot$ + 5 M$_\odot$ 
stars with initial period $\sim 4~P_{\rm zams}$. 
``H'' denotes the end of the main sequence for each star. 
(reproduced from \cite{2001ApJ...552..664N} with permission by the AAS.}
\label{algolA}
\end{figure}

Blue stragglers that are typical for globular cluster conditions and that 
remain in their original binaries could be formed
 via  AS (``slow'' evolution to contact), AG (first contact is on the giant branch), 
AR (rapid contact),  AL and BL (two cases of ``late overtaking'' scenario, 
where the secondary also overfills its Roche lobe\index{Roche lobe}) types of the evolution. 
AR evolution towards blue straggler formation is rare though as it
can occur only with primaries close to 2~M$_\odot$ and for 
most initial mass ratios, the companion does not 
gain enough mass to appear as a blue straggler.
A couple of typical examples of binary mass transfer 
sequences, cases AS and AL,  are shown in Fig.~\ref{algolA} 
(note that the stars that are shown are more massive than 
stars present in \emph{now} globular clusters, but
less massive stars would evolve via scenarios qualitatively similar to the ones shown). 
AS, AL and BL cases usually took place in binaries that had initial mass 
ratios smaller than 2 (at $q> 2$, AD takes place), AR takes place
in binaries with primaries having a mass close to 2~M$_\odot$ and $q>2$.
Most of the binaries that have case A or case B mass transfer 
towards a blue straggler formation have first thermal time-scale mass transfer, 
producing both components being strongly out of their thermal equilibrium. 
The primary after the mass transfer, if it remains a main-sequence star, 
would be out of thermal equilibrium for a long time --- 
a blue straggler in this case could be in a binary where 
the lower-mass companion is out of thermal equilibrium. 

While we have an overall understanding of how blue stragglers 
could be formed via case A and case B mass transfers, 
theory is still far from  predicting exact numbers and binary configurations. 
To test the theory with observations, we need to compare
predicted with observed blue straggler populations. 
Blue stragglers in globular clusters are contaminated by those formed via collisions,
and blue stragglers in a field are not easy to distinguish.  
However, many of the blue stragglers formed via case A and case B mass transfers
are precursors of Algol systems\index{Algol system}, which are much easier to be observationally detected,
and, hence, we can test out the theory on Algol systems.
It was shown that theoretically obtained distributions and numbers of Algols 
still do not match well with observations
(e.g., \cite{2011AA...528A..16V}). The important outcome of 
Algol's studies is that the observed distributions
cannot be matched by conservative evolution, 
though liberal evolution does not reproduce Algols fairly well neither
(also note that less conservative mass transfer is less effective 
in the production of blue stragglers --- by definition, a blue straggler is a star 
more massive than its parent population, i.e. it should have gained enough mass
in a binary system).

\subsection {Late Case B/Case C} 

Here we consider the cases when mass transfer that resulted in 
the formation of a blue straggler has started from a donor with a well 
developed convective envelope.
From the analysis above, binary systems with low-mass giant primaries, 
in conservative evolution, 
are expected to start dynamically unstable mass transfer 
for most case B mass transfers.
In many cases, a binary would survive common envelope evolution\index{common-envelope evolution}, but 
a main sequence companion is not capable of accreting during 
a common envelope phase enough material 
to appear as a blue straggler \cite{1991ApJ...370..709H}.

As discussed above, due to wind mass loss, metal-rich 
globular clusters are more likely to form blue stragglers via conservative case B 
mass transfer than metal-poor globular clusters ---  
in this case, the stripped red giant core would be close to 0.47 M$_\odot$.

Case C mass transfer is expected to be stable in globular 
clusters of all metallicities and will 
lead to the formation of long-orbital-period blue stragglers 
(case C also can be unstable, however, this occur for such mass ratios that a companion
is unlikely to appear as a blue straggler). 

The stated above expectations were based on an adiabatic response of giant donors.
With superadiabatic response taken into account,  blue stragglers 
can be formed via stable case B mass transfers from giant donors  
with smaller core masses. 
Similar effect is achieved by non-conservation 
(though again note the caveat: with non-conservative mass transfer 
it is easier to have stable mass transfer, but harder to accrete 
enough to become a blue straggler).

Liberal mass transfer from red giant donors to non-degenerate 
companions has been discussed in \cite{woods11},
 especially its implications and importance for the  avoidance 
of the dynamically unstable mass transfer from a primary 
(this is the first of the two episodes of the mass transfer that 
eventually forms a double white dwarf system). 
The same channel that produces double white dwarfs also produces 
long-orbital-period blue stragglers at its first leg. 

Forming a blue straggler could also be a precursor for 
hot sub-dwarf formation (for a recent review, see \cite{2013EPJWC..4304001G}). 
Case B channel can produce wide systems with massive white dwarfs, but only
if the case B mass transfer is liberal and/or the giant's response is superadiabatic 
would we be able to find at least some blue stragglers in a relatively 
``narrow''-wide system ($P\la 100$ days) with a low-mass white dwarf  ($\la 0.25$~M$_\odot$).
In this respect, it is interesting to note the recent discovery of a stripped red giant with 
a mass of $0.23\pm0.03$~M$_\odot$, 
in an eclipsing binary system and with an A-type companion \cite{2011MNRAS.418.1156M}, 
though the period of this binary is very short to be a characteristic late case B.
Finally, as was mentioned in \S\ref{sec_howwell}, the mass transfer rates in binaries 
with early giant donors are among the least well understood.

\subsection {Role of Globular Cluster Dynamics on the RLOF}

In globular clusters, dynamical encounters could replace an original 
binary companion by a more massive one
and/or increase the binary's eccentricity. 
Those perturbation, however, do not affect the previously described stability and
rate of the mass transfer once it has started. 
A more interesting deviation from most field binaries is that 
a binary that contains a potential blue straggler 
is much more likely to be a member of a dynamically formed hierarchical triple.

A fraction of dynamically formed hierarchically stable triples, those that have 
inclinations of the orbit $\ga 39^0$, could be affected by the Kozai mechanism\index{Kozai mechanism}
\cite{1962AJ.....67..591K,1997AJ....113.1915I}. Such
Kozai mechanism causes large variations in the eccentricity and inclination 
of the binary orbits and, 
especially if coupled with tidal friction,  could drive the inner binary 
of the triple system to RLOF 
and/or merger \cite{2006Ap&SS.304...75E,2007ApJ...669.1298F}.
If triple induced RLOF leads to a merger, it can form about 10\% 
of blue stragglers in a globular cluster \cite{2008msah.conf..101I},
but whether it will be a merger or a mass transfer requires 
further understanding of the evolution of contact binaries
 \cite{2012JASS...29..145E}.

\begin{acknowledgement}
N. Ivanova acknowledges support from NSERC Discovery and 
Canada Research Chairs programs.
\end{acknowledgement}


\begin{thebibliography}{99}
\providecommand{\url}[1]{{#1}}
\providecommand{\urlprefix}{URL }
\expandafter\ifx\csname urlstyle\endcsname\relax
  \providecommand{\doi}[1]{DOI \discretionary{}{}{}#1}\else
  \providecommand{\doi}{DOI \discretionary{}{}{}\begingroup
  \urlstyle{rm}\Url}\fi


\bibitem{Beg79}
 {Begelman}, M. C.: \mnras \textbf{187}, 237 (1979)

\bibitem{2008ApJS..174..223B}
Belczynski, K.,  {Kalogera}, V.,  {Rasio}, F.~A. et al.:
 \apjs \textbf{174}, 223 (2008)

\bibitem{Dalsgaard2011} 
Christensen-Dalsgaard, J.: in Progress in Solar/Stellar Physics with 
Helio- and Asteroseismology, ASP Conf. \textbf{462}, 503 (2012)

\bibitem{dar79}
{Darwin}, G. H.: Proc. Roy. Soc. London \textbf{29}, 168 (1879)

\bibitem{2003ApJ...598.1217D}
{Deloye}, C. J.,~{Bildsten}, L.: \apj \textbf{598}, 1217 (2003)

\bibitem{dem91}
{Demircan}, O., {Kahraman}, G.: \apss \textbf{181}, 313 (1991)

\bibitem{1971MNRAS.151..351E}
 {Eggleton}, P. P.: \mnras \textbf{151}, 351 (1971)

\bibitem{1972MNRAS.156..361E}
{Eggleton}, P. P.: \mnras \textbf{156}, 361 (1972)

\bibitem{1973MNRAS.163..279E}
{Eggleton}, P. P.: \mnras \textbf{163}, 279 (1973)

\bibitem{Eggl83}
{Eggleton}, P. P.: \apj \textbf{268}, 368 (1983)

\bibitem{2000NewAR..44..111E}
{Eggleton}, P. P.: New Astronomy Reviews \textbf{44}, 111 (2000)

\bibitem{Eggl06}
{Eggleton}, P. P.: Evolutionary Processes in Binary and Multiple Stars, Cambridge University Press
  (2006)

\bibitem{2012JASS...29..145E}
{Eggleton}, P. P.: Journal of Astronomy and Space Sciences \textbf{29}, 145
  (2012)

\bibitem{1973AA....23..325E}
{Eggleton}, P. P., {Faulkner}, J.,~{Flannery}, B. P.: \aap \textbf{23}, 325 (1973)

\bibitem{2006Ap&SS.304...75E}
{Eggleton}, P. P., {Kisseleva-Eggleton}, L.: \apss \textbf{304}, 75 (2006)

\bibitem{2011ascl.soft07008E}
{Eggleton}, P. P., {Tout}, C., {Pols}, O. R., et al.:
{STARS: A Stellar Evolution Code}, Astrophysics Source Code Library (2011)

\bibitem{2007ApJ...669.1298F}
{Fabrycky}, D., {Tremaine}, S.: \apj \textbf{669}, 1298 (2007)

\bibitem{Ge2010}
H.~{Ge}, M. S. {Hjellming}, R. F. {Webbink}, X.~{Chen}, Z.~{Han}: \apj
  \textbf{717}, 724 (2010)

\bibitem{2013EPJWC..4304001G}
 Geier, S.: European Physical Journal Web of Conferences \textbf{43}, 4001 (2013)
      
\bibitem{2008AA...488.1007G}
~{Glebbeek}, E.,  {Pols}, O. R., {Hurley}, J. R.: \aap \textbf{488}, 1007 (2008)

\bibitem{heinke2013}
 {Heinke}, C. O., {Ivanova}, N.,  {Engel}, M. C., et al.:
  \apj \textbf{768}, 184 (2013)

\bibitem{Hje87}
 {Hjellming}, M. S., {Webbink}, R. F.: \apj \textbf{318}, 794 (1987)

\bibitem{1991ApJ...370..709H}
 {Hjellming}, M. S., {Taam}, R. E.: \apj \textbf{370}, 709 (1991)

\bibitem{iva2002MNRAS.329..897H}
 Hurley, J.~R., {Tout}, C.~A.,  {Pols}, O.~R.: \mnras \textbf{329}, 897 (2002)
      
\bibitem{Hut80}
{Hut}, P.: \aap \textbf{92}, 167 (1980)

\bibitem{2007AA...465..953I}
 {in't Zand}, J. J. M., {Jonker}, P. G. ,  {Markwardt}, C.B.: \aap \textbf{465}, 953
  (2007)

\bibitem{1997AJ....113.1915I}
 {Innanen}, K. A., {Zheng}, J. Q., {Mikkola}, S.,  {Valtonen}, M.~J.: \aj \textbf{113},
  1915 (1997)

\bibitem{Iva03}
{Ivanova}, N.: DPhil Thesis  (2003)

\bibitem{2008msah.conf..101I}
{Ivanova}, N.: in Multiple Stars Across the H-R Diagram, ESO Symp., Springer-Verlag, p. 101 (2008)

\bibitem{ips02}
{Ivanova}, N., {Podsiadlowski}, P., {Spruit}, H.: \mnras \textbf{334}, 819 (2002)

\bibitem{Iva04}
{Ivanova}, N.,  {Taam}, R.E.: \apj \textbf{601}, 1058 (2004)

\bibitem{iva_ce}
{Ivanova}, N., {Justham}, S., {Chen}, X., et al.:
\aapr \textbf{21}, 59  (2013)

\bibitem{kipwei67}
{Kippenhahn}, R., {Weigert}, A.: \zap \textbf{65}, 251 (1967)

\bibitem{Kip67}
{Kippenhahn}, R., {Weigert}, A., {Hofmeister}, E.: Mth. Comp. Phys. \textbf{7},
  129 (1967)

\bibitem{1962AJ.....67..591K}
{Kozai}, Y.: \aj \textbf{67}, 591 (1962)

\bibitem{2008IAUS..250..167L}
  {Langer}, N., {Cantiello}, M.,  {Yoon},  S.-C., et al.: 
	in IAU Symposium 250, pp-167 (2008)

\bibitem{ivalau70}
{Lauterborn}, D.: \aap \textbf{7}, 150 (1970)

\bibitem{2002ApJ...568..939L}
 {Lombardi}, Jr., J. C., {Warren},  J.S., {Rasio}, F. A. , {Sills},  A.,  {Warren}, A.~R.:
  \apj \textbf{568}, 939 (2002)

\bibitem{2011ApJ...737...49L}
{Lombardi}, Jr., J. C. , {Holtzman}, W., {Dooley}, K. L. ,~{Gearity},  K., {Kalogera}, V., {Rasio}, F. A.: \apj \textbf{737}, 49 (2011)

\bibitem{1976ApJ...207L..53L}
 {Lubow}, S. H.,  {Shu}, F. H.: \apjl \textbf{207}, L53 (1976)

\bibitem{2011MNRAS.418.1156M}
{Maxted}, P. F. L.,  {Anderson}, D. .,  {Burleigh}, M. R., et al.: 
  \mnras \textbf{418}, 1156 (2011)

\bibitem{2001ApJ...552..664N}
 {Nelson}, C. A.,  {Eggleton}, P. P.: \apj \textbf{552}, 664 (2001)

\bibitem{Nom79}
{Nomoto}, K., {Nariai}, K., {Sugimoto}, D.: \pasj \textbf{31}, 287 (1979)

\bibitem{1981A&A...102...17P}
Packet, W.: \aap \textbf{102}, 17 (1981)

\bibitem{2012ApJ...760...90P}
 {Passy}, J. C., {Herwig}, F., {Paxton}, B.: \apj \textbf{760}, 90 (2012)

\bibitem{2012ApJ...744...52P}
 {Passy}, J.C., {De Marco}, O.,  {Fryer}, C.L., et al.:
   \apj \textbf{744},
  52 (2012)

\bibitem{pavl_casca}
{Pavlovskii}, K., {Ivanova}, N.:  CASCA meeting (2013)

\bibitem{2013IAUS..291..468P}
{Pavlovskii}, K., {Ivanova}, N.: in IAU Symposium 291, 468 (2013)

\bibitem{2013IAUS..290..293P}
{Pavlovskii}, K., {Ivanova}, N.: in IAU Symposium, 290, 293 (2013)

\bibitem{MESA1}
{Paxton}, B.,~{Bildsten},  L., {Dotter}, A., et al.:
  \apjs \textbf{192}, 3 (2011)

\bibitem{MESA2}
Paxton, B., Cantiello, 
M., Arras, P., et al.: \apjs \textbf{208}, 4  (2013)

\bibitem{podsi02}
{Podsiadlowski}, P., {Rappaport}, S.,  {Pfahl}, E. D.: \apj \textbf{565}, 1107
  (2002)

\bibitem{1995MNRAS.274..964P}
 {Pols}, O. R., {Tout}, C. A. ,  {Eggleton}, P. P., {Han}, Z.: \mnras \textbf{274}, 964
  (1995)

\bibitem{2012MNRAS.427.1702R}
{Raymer}, E.: \mnras \textbf{427}, 1702 (2012)

\bibitem{2012ApJ...746...74R}
 {Ricker}, P. M., {Taam}, R. E.: \apj \textbf{746}, 74 (2012)

\bibitem{1980ApJ...239..937S}
 Shu, F.~H.,   {Lubow}, S.~H., {Anderson}, L.: 
{\apj}, \textbf{239}, 937 (1980)
      
\bibitem{Sob97}
 {Soberman}, G. E., {Phinney}, , E. S.,  {van den Heuvel}, E. P. J.: \aap \textbf{327},
  620 (1997)

\bibitem{2008MNRAS.389.1828S}
 Stancliffe, R.~J.,  Glebbeek, E.:
\mnras \textbf{389}, 1828 (2008)
      
\bibitem{Tout97}
{Tout}, C.~A.,  {Aarseth}, S.~J.,  {Pols}, O. R., {Eggleton}, P. P.: \mnras \textbf{291},
  732 (1997)
  
\bibitem{2008AA...487.1129V}
{van Rensbergen}, W.,  {De Greve}, J.P., {De Loore}, C., {Mennekens}, N.: \aap
  \textbf{487}, 1129 (2008)

\bibitem{2011AA...528A..16V}
{van Rensbergen},  W., {De Greve},J.P., ~{Mennekens},  N.,
	{Jansen}, K. , de Loore, C.: 
 {\aap},\textbf{528}, A16 (2011)
      
\bibitem{webbink1985}
{Webbink}, R.F.: in Stellar evolution and binaries, Cambridge University Press, p.~39 (1985)

\bibitem{woodsmt11}
{Woods}, T.E.,  {Ivanova}, N.: \apjl \textbf{739}, L48 (2011)

\bibitem{woods11}
{Woods}, T.E., {Ivanova}, N.~, {van der Sluys}, M.V. , {Chaichenets}, S.~: \apj
  \textbf{744}, 12 (2012)

\bibitem{1997MNRAS.292..321Y} Yukawa, H.,  Boffin, H.~M.~J., Matsuda, T. : \mnras \textbf{292}, 321 (1997)

\end{thebibliography}

%

\backmatter
\printindex


\end{document}